\documentclass{PoS}

\title{Lattice QCD Study for Gluon Propagator and \\
Gluon Spectral Function}

\ShortTitle{Lattice QCD Study for Gluon Propagator 
and Gluon Spectral Function}

\author{\speaker{Hideo Suganuma}, Takumi Iritani, Arata Yamamoto \\
        Department of Physics, Kyoto University, 
Kitashirakawaoiwake, Sakyo, Kyoto 606-8502, Japan\\
        E-mail: \email{suganuma@scphys.kyoto-u.ac.jp}
}

\author{Hideaki Iida \\
The Institute of Physical and Chemical Research (RIKEN), 
Wako, Saitama 351-0198, Japan}

\abstract{
We study the gluon propagator in the Landau gauge in 
SU(3) lattice QCD at $\beta$=5.7, 5.8 and 6.0 at the quenched level. 
The Euclidean Landau-gauge gluon propagator 
$D(r)\equiv D_{\mu\mu}^{aa}(x)/24$ 
is found to be well described by four-dimensional Yukawa-type function 
$e^{-mr}/r$ in the infrared and intermediate region of 
$r \equiv (x_\mu x_\mu)^{1/2}$ = 0.1 $\sim$ 1.0fm. 
The infrared effective gluon mass is obtained as $m \simeq$ 600MeV. 
Associated with the 4D Yukawa-type gluon propagator, 
we derive analytical expressions for 
the zero-spatial-momentum propagator $D_0(t)$, 
the effective mass $M_{\rm eff}(t)$, and the spectral function 
$\rho(\omega)$ of the gluon field. 
Remarkably, the obtained gluon spectral function   
$\rho(\omega)$ is almost negative definite, 
except for a positive $\delta$-functional peak at $\omega=m$. 
Since the Yukawa-type propagation indicates 
a three-dimensional space-time character, 
we consider a hypothesis of an effective dimensional reduction 
by generalized Parisi-Sourlas mechanism 
in a stochastic color-magnetic vacuum of infrared QCD. 
}

\FullConference{Lattice Field Theories 2010\\
                 June 12-19, 2010\\
                 Tanka, Sardinia, Italy}

\begin{document}

\section{Introduction}

The analysis of gluon properties is an important key point 
to clarify the nonperturbative aspects of QCD \cite{ISI09,YS0809,C82}.
In particular, the gluon propagator, {\it i.e.}, 
the two-point Green function is one of the most basic quantities in QCD, 
and has been investigated with much interests 
\cite{ISI09,HUGE,AS99}.
Dynamical gluon-mass generation is also an important subject 
related to the infrared gluon propagation. 
While gluons are perturbatively massless, 
they are conjectured to acquire a large effective mass 
as the self-energy through their self-interaction 
in a nonperturbative manner \cite{C82,ABP08}.
Actually, glueballs, color-singlet bound states of gluons, 
are theoretically predicted to be fairly massive, 
{\it e.g.}, about 1.5GeV for the lowest $0^{++}$ and about 2GeV 
for the lowest $2^{++}$, 
in lattice QCD calculations \cite{ISM02}. 

For the direct investigation of the gluon field, gauge fixing is to be done. 
Among gauges, the Landau gauge is one of the most popular gauges in QCD, 
and it keeps Lorentz covariance and global ${\rm SU}(N_c)$ symmetry.
In Euclidean QCD, the Landau gauge has a global definition 
to minimize the global quantity 
$
R \equiv \int d^4 x \ {\rm Tr} \{A_\mu(x) A_\mu(x)\} 
= \frac{1}{2} \int d^4 x A_\mu^a(x) A_\mu^a(x)
$
by gauge transformation.
The local condition $\partial_\mu A_\mu(x) = 0$ 
is derived from the minimization of $R$.
The global quantity $R$ can be regarded as 
``total amount of the gluon-field fluctuation'' in Euclidean space-time. 
In the global definition, 
the Landau gauge has a clear physical interpretation 
that it maximally suppresses artificial gauge-field fluctuations 
relating to gauge degrees of freedom \cite{ISI09}.

In lattice QCD, the Landau gauge is defined by the maximization of 
$
R_{\rm latt} \equiv \sum_x \sum_\mu {\rm Re} {\rm Tr} U_\mu(x),
$
with the link-variable $U_\mu(x) \equiv e^{iagA_\mu(x)}$ 
($a$: lattice spacing, $g$: QCD gauge coupling).
The gluon field is defined as 
$
A_\mu(x) \equiv \frac{1}{2iag} 
\{ U_\mu(x) - U_\mu^\dagger(x) \} - ({\rm trace \ part}).
$
In the Landau gauge, the minimization of gluon-field fluctuations 
justifies the expansion by small lattice spacing $a$. 
In Euclidean metric, 
the gluon propagator is defined by the two-point function as  
$
D_{\mu\nu}^{ab}(x-y) \equiv \langle A_\mu^a(x) A_\nu^b(y) \rangle.
$
Here, owing to the symmetries and the transverse property, 
the color and Lorentz structure of the gluon propagator 
is uniquely determined in the Landau gauge.

In this paper, 
using SU(3) lattice QCD Monte Carlo calculations, 
we study the functional form of 
the Landau-gauge gluon propagator, 
$
D(r) \equiv \frac{1}{3 (N_c^2-1) } D_{\mu\mu}^{aa}(x)
     = \frac{1}{3 (N_c^2-1) } 
\langle A_\mu^a(x)A_\mu^a(0)\rangle,
$
as a function of 4D Euclidean distance 
$r \equiv (x_\mu x_\mu)^{1/2}$.
We mainly deal with the coordinate-space propagator $D(r)$ 
for the infrared and intermediate region of $r=0.1 \sim 1.0$fm, 
which is relevant for quark-hadron physics.
Based on the obtained function form of the gluon propagator, 
we aim at a nonperturbative description of gluon properties,

\vspace{-0.1cm}

\section{Functional form of Landau-gauge gluon propagator}

\vspace{-0.1cm}

The ${\rm SU}(3)$ lattice QCD Monte Carlo calculations 
are performed at the quenched level 
using the standard plaquette action with 
$\beta \equiv 2N_c/g^2$=5.7, 5.8, and 6.0, 
on the lattice size of 
$16^3 \times 32$, $20^3 \times 32$, and $32^4$, respectively.
The lattice spacing $a$ is found to be $a = 0.186, 0.152$, and $0.104$fm, 
at $\beta$ = 5.7, 5.8, and 6.0, respectively, 
when the scale is determined so as to reproduce the string tension as 
$\sqrt{\sigma} = 427$MeV 
from the static Q$\bar {\rm Q}$ potential \cite{STI04}.
Here, we choose the renormalization scale at $\mu=4{\rm GeV}$ 
for $\beta=6.0$, and make corresponding rescaling for $\beta$=5.7 and 5.8 
\cite{ISI09}.

Figure 1(a) and (b) show the coordinate-space gluon propagator $D(r)$ and 
the momentum-space gluon propagator 
$\tilde D(p^2)\equiv \int d^4x \ e^{ip \cdot x} D(r)$, respectively.
Our lattice QCD result of $\tilde D(p^2)$ is 
consistent with that obtained in previous lattice studies,
although recent huge-volume lattice studies \cite{HUGE} 
indicate a suppression of the gluon propagator 
in the Deep-IR region ($p < 0.5$GeV).

\begin{figure}[t]
\begin{center}
\includegraphics[scale=0.9]{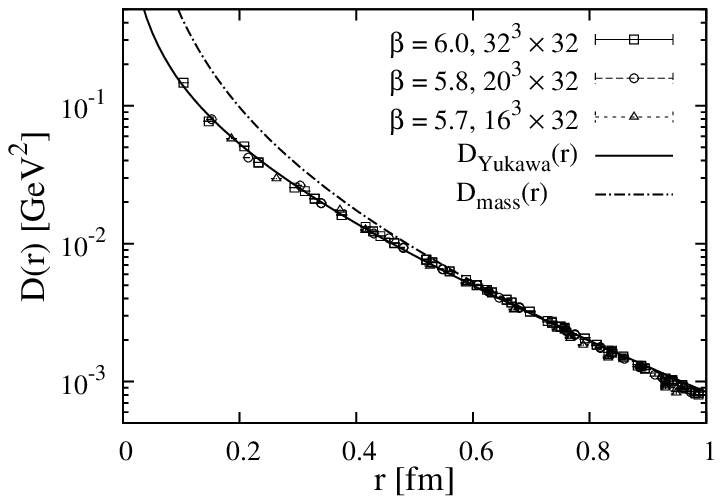}
\includegraphics[scale=0.9]{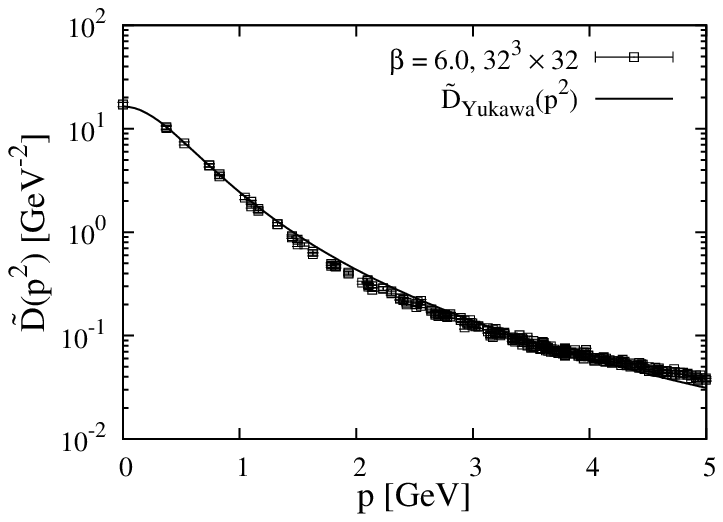}
\caption{
(a) Lattice QCD results (symbols) of the Landau-gauge gluon propagator 
$D(r)\equiv D_{\mu\mu}^{aa}(x)/24$ 
in coordinate space, 
and 4D Yukawa-type function 
$D_{\rm Yukawa}(r)=Ame^{-mr}/r$~ (solid line) 
with $m=0.624$GeV and $A=0.162$.
The dash-dotted line denotes 
a typical example of the massive-vector propagator $D_{\rm mass}(r)$.
(b) The Landau-gauge gluon propagator 
$\tilde{D}(p^2)$ in the momentum space of 
$p_\mu=\frac{2}{a}\sin(\frac{\pi n_\mu}{L_\mu})$.
The solid line denotes 4D Fourier-transformed 
Yukawa-type propagator, {\it i.e.}, 
$\tilde{D}_{\rm Yukawa}(p^2) = 4\pi^2 Am (p^2+m^2)^{-3/2}$.
}
\end{center}
\vspace{-0.2cm}
\end{figure}

We find that the lattice gluon propagator $D(r)$ cannot be described by 
the free massive Euclidean propagator 
$
D_{\rm mass}(r) = \int \frac{d^4p}{(2\pi)^4} e^{-ip\cdot x}\frac{1}{p^2+m^2} 
= \frac{1}{4\pi^2} \frac{m}{r} K_1(mr)
$
($K_\nu(z)$: modified Bessel function) \cite{AS99} 
in the whole region of $r = 0.1 \sim 1.0$fm, as shown in Fig.1(a).

By the functional-form analysis, 
we find that 
the Landau-gauge gluon propagator $D(r)$ in the coordinate space 
is well described by the 4D Yukawa-type function \cite{ISI09}
\begin{equation}
D(r) \equiv \frac{1}{24} D_{\mu\mu}^{aa}(r)= Am \frac{e^{-mr}}{r},
\end{equation}
with $m=0.624(8)$GeV and $A=0.162(2)$ 
in the range of $r=0.1 \sim 1.0$fm, 
as shown in Fig.1(a).
The gluon propagator $\tilde{D}(p^2)$ in the momentum space 
is also well described by 4D Fourier-transformed Yukawa-type function 
as 
$
\tilde{D}(p^2) = \frac{1}{24} \tilde{D}_{\mu\mu}^{aa}(p^2)
= \frac{4\pi^2Am}{(p^2+m^2)^{3/2}}
$
for $0.5{\rm GeV} \le p \le 3{\rm GeV}$ \cite{ISI09}.

\section{\label{sec:effmassanalytic} 
Analytical applications}

In this section, as applications of the Yukawa-type gluon propagator, 
we derive analytical expressions for 
the zero-spatial-momentum propagator $D_0(t)$, 
the effective mass $M_{\rm eff}(t)$, 
and the spectral function $\rho(\omega)$ of the gluon field \cite{ISI09}. 
All the derivations can be analytically performed, 
starting from the Yukawa-type gluon propagator $D_{\rm Yukawa}(r)$.
Although the real gluon propagator has some deviation 
from the Yukawa-type in UV region, 
this method is found to be workable to reproduce lattice QCD results, 
as shown below.

\subsection{Zero-spatial-momentum propagator of gluons}

First, we consider zero-momentum gluon propagator 
$
D_0(t) \equiv \frac{1}{24} \sum_{\vec{x}} 
\langle A_\mu^a(\vec{x},t) A_\mu^a(\vec{0},0)\rangle
= \sum_{\vec{x}} D(r),
$
where $r = \sqrt{\vec{x}^2+t^2}$ is the 4D Euclidean distance.
For the simple argument, we here deal with the continuum formalism 
with infinite space-time. 
Starting from the Yukawa-type gluon propagator $D_{\rm Yukawa}(r)$, 
we derive the zero-spatial-momentum propagator as \cite{ISI09}
\begin{equation}
D_0(t) 
=\int d^3 x \ D_{\rm Yukawa}(\sqrt{\vec{x}^2+t^2})
=4\pi Am \int_0^{\infty} dx 
\frac{x^2}{\sqrt{x^2+t^2}} e^{-m\sqrt{x^2+t^2}}
=4\pi A t K_1(mt).
\label{eq:contZMP}
\end{equation}
In Fig.2(a), we show the theoretical curve of $D_0(t)$ 
in Eq.(\ref{eq:contZMP}) with $m$=0.624GeV and $A$=0.162, 
together with the lattice QCD result of $D_0(t)$ in the Landau gauge. 
For the actual comparison with the lattice data, 
we take account of the temporal periodicity \cite{ISI09}.
The lattice QCD data are found to be well described by  
the theoretical curve, associated with the Yukawa-type gluon propagator. 

\begin{figure}[t]
\begin{center}
\includegraphics[scale=0.95]{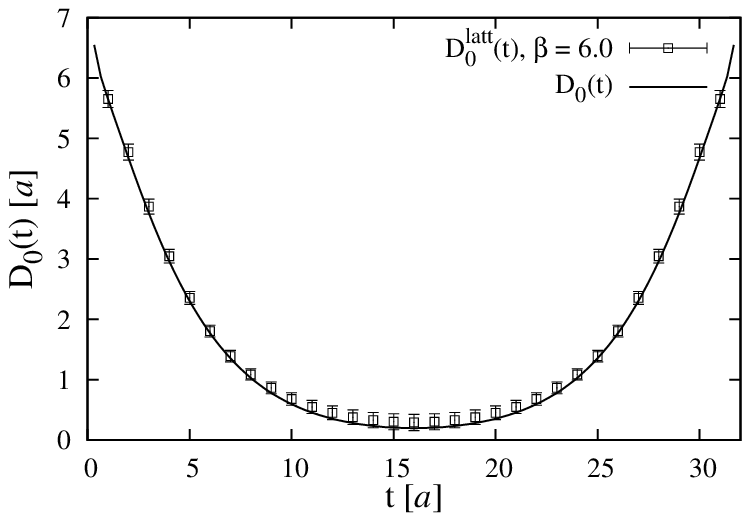}
\includegraphics[scale=0.95]{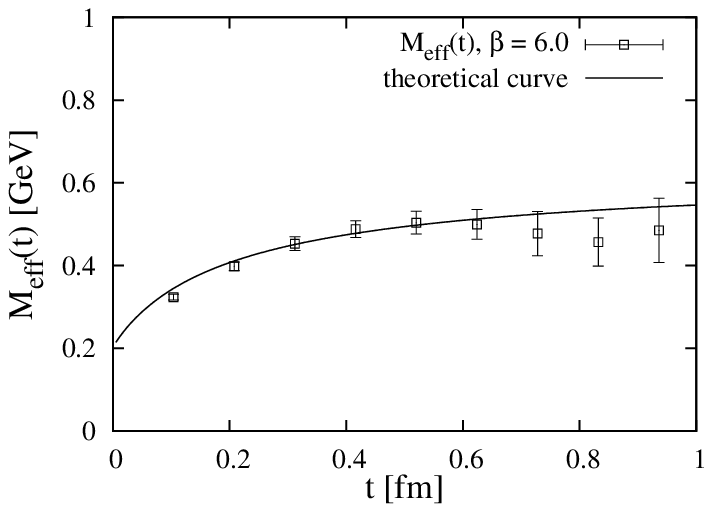}
\caption{
(a) The zero-spatial-momentum propagator $D_0(t)$ of gluons 
in the Landau gauge.
(b) The effective mass $M_{\rm eff}(t)$ of gluons in the Landau gauge.
The symbols are the lattice QCD data on $32^4$ at $\beta = 6.0$,
and the solid line is the theoretical curve derived 
from 4D Yukawa-type propagator with $m$=0.624GeV and $A$=0.162.
}
\end{center}
\vspace{-0.5cm}
\end{figure}

\subsection{Effective mass of gluons}

Second, we investigate the effective mass $M_{\rm eff}(t)$ of gluons.
The effective mass plot is often used for hadrons 
as a standard mass measurement in lattice QCD. 
For the simple notation, we use the lattice unit of $a=1$ in this subsection.
In the case of large temporal size, 
the effective mass is defined as 
$
M_{\rm eff}(t) = \ln \{D_0(t)/D_0(t+1)\}.
$

In Fig.2(b), we show the lattice result of $M_{\rm eff}(t)$, 
where we take account of the temporal periodicity.
The effective gluon mass exhibits a significant scale-dependence, 
and it takes a small value at short distances.
Quantitatively, the effective gluon mass is estimated 
to be about $400 \sim 600$MeV 
in the infrared region of about 1fm \cite{ISI09}.
This value seems consistent with the gluon mass suggested 
by Cornwall \cite{C82}, from a systematic analysis of 
nonperturbative QCD phenomena.

Now, we consider the consequence of 4D Yukawa-type propagator 
$D_{\rm Yukawa}(r)$ of gluons.
For simplicity, we here treat the three-dimensional space 
as a continuous infinite-volume space, 
while the temporal variable $t$ is discrete.
We obtain an analytical expression of 
the effective mass \cite{ISI09}, 
\begin{equation}
M_{\rm eff}(t) = \ln \frac{D_0(t)}{D_0(t+1)}
=\ln \frac{t K_1(mt)}{(t+1)K_1(m(t+1))},
\label{eq:EMGYukawa}
\end{equation}
when the temporal periodicity can be neglected.
In Fig.2(b), we add by the solid line 
the theoretical curve of $M_{\rm eff}(t)$ in 
Eq.(\ref{eq:EMGYukawa}) with $m$=0.624GeV.
The lattice QCD data of $M_{\rm eff}(t)$ are found to be well described by  
the theoretical curve derived from the Yukawa-type gluon propagator. 
From the asymptotic form $K_1(z) \propto z^{-1/2}e^{-z}$, 
the effective mass of gluons is approximated as \cite{ISI09} 
\begin{equation}
M_{\rm eff}(t) \simeq 
m - \frac{1}{2} \ln \big( 1 + \frac{1}{t} \big)
\simeq  m - \frac{1}{2t}
\qquad ({\rm for} \ \ {\rm large} \ \ t).
\end{equation} 
This functional form indicates that 
$M_{\rm eff}(t)$ is an increasing function and approaches $m$ 
from below, as $t$ increases.
Then, the mass parameter $m \simeq$ 600MeV 
in the Yukawa-type gluon propagator has a definite physical meaning 
of the effective gluon mass in the infrared region.

Note that the simple analytical expression  
reproduces the anomalous ``increasing behavior'' of 
the effective mass $M_{\rm eff}(t)$ of gluons.
Thus, this framework with the Yukawa-type gluon propagator  
gives an analytical and quantitative method, and 
is found to well reproduce lattice QCD results.

\subsection{Spectral function of gluons in the Landau gauge}

As a general argument, 
an increasing behavior of the effective mass $M_{\rm eff}(t)$ 
means that the spectral function 
is not positive-definite \cite{ISI09}.
More precisely, the increasing property of $M_{\rm eff}(t)$ can be realized, 
only when there is some suitable coexistence of positive- and negative-value 
regions in the spectral function $\rho(\omega)$ \cite{ISI09}.
However, the functional form of the spectral function 
of the gluon field is not yet known.

The relation between the spectral function $\rho(\omega)$ and 
the zero-spatial-momentum propagator $D_0(t)$ 
is given by the Laplace transformation, 
$
D_0(t) = \int_0^\infty d\omega \ \rho(\omega) \ e^{-\omega t}.
$
When the spectral function is given by a $\delta$-function 
such as $\rho(\omega) \sim \delta(\omega-\omega_0)$, 
which corresponds to a single mass spectrum, 
one finds a familiar relation of $D_0(t) \sim e^{-\omega_0t}$.
For the physical state, the spectral function $\rho(\omega)$ gives 
a probability factor, and is non-negative definite 
in the whole region of $\omega$.
This property is related to the unitarity of the S-matrix.

From the analytical expression of the zero-spatial-momentum propagator  
$D_0(t)=4\pi A t K_1(mt)$, 
we can derive the spectral function $\rho(\omega)$ of the gluon field, 
associated with the Yukawa-type gluon propagator \cite{ISI09}.
For simplicity, 
we take continuum formalism with infinite space-time.
Using the inverse Laplace transformation of the modified Bessel function, 
we derive the spectral function $\rho(\omega)$ 
of the gluon field as \cite{ISI09} 
\begin{equation}
\rho(\omega)
= -\frac{4\pi A m}{(\omega^2-m^2)^{3/2}}\theta(\omega-m-\varepsilon)
+\frac{4\pi A/\sqrt{2m}}{(\omega-m)^{1/2}} \delta(\omega-m-\varepsilon),
\label{eq:SFYukawa}
\end{equation}
with an infinitesimal positive $\varepsilon$, 
which is introduced for a regularization.
Here, $m \simeq$ 600MeV is the mass parameter in the Yukawa-type function  
for the Landau-gauge gluon propagator.
The first term expresses a negative continuum spectrum, and 
the second term a $\delta$-functional peak with 
the residue including a positive infinite factor as $\varepsilon^{-1/2}$ 
at $\omega=m+\varepsilon$.

\begin{figure}[t]
\begin{center}
\includegraphics[scale=1]{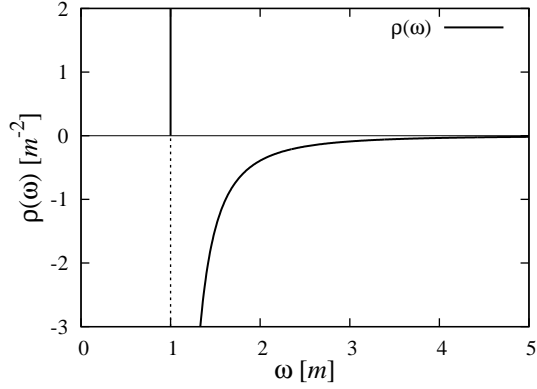}
\caption{\label{fig:Spectral}
The spectral function $\rho(\omega)$ of the gluon field, 
associated with the Yukawa-type propagator.
The unit is normalized by the mass parameter $m \simeq$ 600MeV. 
$\rho(\omega)$ shows anomalous behaviors: 
it has a positive $\delta$-functional peak 
with the residue of $+\infty$ at $\omega = m$, 
and takes negative values for all the region of $\omega > m$.
}
\end{center}
\vspace{-0.5cm}
\end{figure}

We show in Fig.3 the spectral function $\rho(\omega)$ of the gluon field.
As a remarkable fact, the obtained gluon spectral function $\rho(\omega)$ 
is negative-definite for all the region of $\omega > m$, 
except for the positive $\delta$-functional peak at $\omega=m$.
The negative property of the spectral function in coexistence 
with the positive peak leads to the anomalous ``increasing behavior'' 
of the effective mass $M_{\rm eff}(t)$ of gluons \cite{ISI09}.
Actually, Eq.(\ref{eq:SFYukawa}) leads to 
Eq.(\ref{eq:EMGYukawa}), which well describes 
the lattice result of the effective mass $M_{\rm eff}(t)$, 
as shown in Fig.2(b).

We note that the gluon spectral function $\rho(\omega)$ is divergent 
at $\omega=m+\varepsilon$, 
and there are {\it two} divergence structures: 
a $\delta$-functional peak with a positive infinite residue 
and a negative wider power-damping peak.
On finite-volume lattices, these singularities are to be smeared, 
and $\rho(\omega)$ is expected to take a finite value everywhere on $\omega$. 
On the lattice, the spectral function $\rho(\omega)$ 
is conjectured to include a narrow ``positive-valued peak'' 
stemming from the $\delta$-function 
in the vicinity of $\omega=m \ (+\varepsilon)$ 
and a wider ``negative-valued peak'' near $\omega \simeq m$ 
in the region of $\omega > m$ \cite{ISI09}.

In this way, the Yukawa-type gluon propagator indicates 
an extremely anomalous spectral function 
of the gluon field in the Landau gauge. 
The obtained gluon spectral function $\rho(\omega)$ 
is negative almost everywhere, and 
includes a complicated divergence structure 
near the ``anomalous threshold'', $\omega=m\ (+\varepsilon)$. 
Thus, this framework with the Yukawa-type gluon propagator 
gives an analytical and concrete expression 
for the gluon spectral function $\rho(\omega)$.

\vspace{-0.15cm}

\section{Effective dimensional reduction in gluonic vacuum 
by Parisi-Sourlas mechanism}

We discuss the Yukawa-type gluon propagation and 
a possible dimensional reduction 
due to the stochastic behavior of the gluon field 
in the infrared region \cite{ISI09}. 
As shown before, the Landau-gauge gluon propagator is well described 
by the Yukawa function in {\it four}-dimensional Euclidean space-time. 
However, the Yukawa function $e^{-mr}/r$ is a natural form 
in {\it three}-dimensional Euclidean space-time, 
since it is obtained by the three-dimensional Fourier transformation of 
the ordinary massive propagator $(p^2+m^2)^{-1}$. 
In fact, the Yukawa-type propagator has a ``three-dimensional'' property.
In this sense, as an interesting possibility, 
we propose to interpret this Yukawa-type behavior of the gluon propagation 
as an ``effective reduction of the space-time dimension''.

Such a ``dimensional reduction'' sometimes occurs in stochastic systems, 
as Parisi and Sourlas pointed out 
for the spin system in a random magnetic field \cite{PS79}.
In fact, on the infrared dominant diagrams, the $D$-dimensional system 
coupled to the Gaussian-random external field 
is equivalent to the $(D-2)$-dimensional system without the external field,
due to a hidden SUSY structure.  

We note that the gluon propagation in the QCD vacuum resembles 
the situation of the system coupled to the stochastic external field.
Actually, as is indicated by a large positive value of the gluon condensate 
$\langle G_{\mu\nu}^aG^{\mu\nu}_a\rangle
=2({\bf H}_a^2-{\bf E}_a^2) >0$ in the Minkowski space,  
the QCD vacuum is filled with a strong color-magnetic field \cite{S77}, 
which can contribute spontaneous chiral-symmetry breaking \cite{ST9193}, 
and the color-magnetic field is considered to be highly random 
at the infrared scale.
Since gluons interact with each other, 
the propagating gluon is violently scattered 
by the other gluons in the 
randomly-oriented color-magnetic fields 
of the infrared QCD vacuum, as shown in Fig.4.

\begin{figure}[t]
\begin{center}
\includegraphics[scale=0.83]{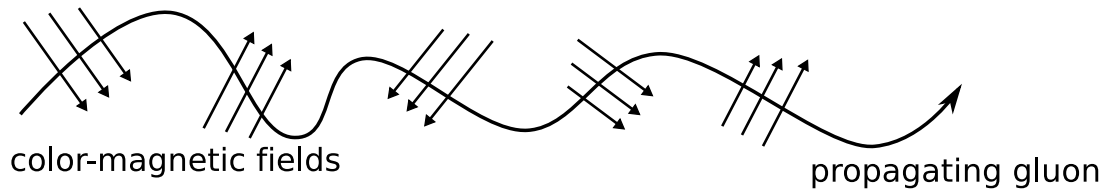}
\caption{
A schematic figure for a propagating gluon. 
The QCD vacuum is filled with color-magnetic fields 
which are stochastic at an infrared scale, 
and the gluon propagates in the random color-magnetic fields.
}
\end{center}
\vspace{-0.6cm}
\end{figure}

Actually at the infrared scale, 
the gluon field shows a strong randomness due to the strong interaction, 
and this infrared strong randomness 
is considered to be responsible for color confinement, 
as is indicated in strong-coupling lattice QCD. 
Even after the removal of fake gauge degrees of freedom by gauge fixing, 
the gluon field exhibits a strong randomness 
accompanying a quite large fluctuation at the infrared scale.

As a generalization of the Parisi-Sourlas mechanism, 
we conjecture that the infrared structure of a theory 
in the presence of quasi-random external fields 
in higher-dimensional space-time has a similarity to 
the theory without the external field 
in lower-dimensional space-time \cite{ISI09}.
From this point of view, the Yukawa-type propagation of gluons 
may indicate an ``effective reduction of space-time dimension'' by one, 
reflecting the interaction between the propagating gluon 
and the other gluons in randomly-oriented color-magnetic fields
in the infrared QCD vacuum.

In any case, it is an interesting and important subject to clarify 
the nonperturbative QCD vacuum structure in terms of  
gluonic properties \cite{YS0809} including the gluon propagation \cite{ISI09}. 

\vspace{-0.2cm}

\section*{Acknowledgements}
\vspace{-0.2cm}
H.S. is supported by Grant-in-Aid for Scientific Research 
[(C) No.~19540287, Priority Areas ``New Hadrons'' (E01:21105006)].
The lattice calculations are done on SX-8R at Osaka University.

\end{document}